\documentclass[aps,pra,showpacs,superscriptaddress,amssymb,twocolumn]{revtex4-1}
\usepackage{graphicx}
\usepackage{appendix} 
\usepackage{hyperref}
\usepackage{bm,amsmath}
\usepackage{dcolumn}

\begin{document}

\rightline{\tt Physical Review A {\bf 92}, 032306 (2015)}

\vspace{0.2in}

\title{Three-step implementation of any $n\!\times\!n$ unitary with a complete graph of $n$ qubits}

\author{Amara Katabarwa}
\email{akataba@uga.edu}
\affiliation{Department of Physics and Astronomy, University of Georgia, Athens, Georgia 30602, USA}
\author{Michael R. Geller}
\email{mgeller@uga.edu}
\affiliation{Department of Physics and Astronomy, University of Georgia, Athens, Georgia 30602, USA}

\date{\today}

\begin{abstract}
Quantum computation with a complete graph of superconducting qubits has been recently proposed, and applications to amplitude amplification, phase estimation, and the simulation of realistic atomic collisions given $[{\rm Phys.~Rev.~A} \ {\bf 91}, 062309 \, (2015)]$. This single-excitation subspace (SES) approach does not require error correction and is practical now. Previously it was shown how to implement symmetric  $n\!\times\!n$ unitaries in a single step, but not general unitaries. Here we show that any element in the unitary group ${\rm U}(n)$ can be executed in no more than three steps, for any $n$. This enables the implementation of highly complex operations in {\it constant} time, and in some cases even allows for the compilation of an entire algorithm down to only three operations. Using this protocol we show how to prepare any pure state of an SES chip in three steps, and also how to compute, for a given SES state $\rho,$ the expectation value ${\rm Tr} (\rho O)$ of any $n\!\times\!n$ Hermitian observable $O$ in a constant number of steps.
\end{abstract}


\maketitle

\section{PRETHRESHOLD QUANTUM COMPUTATION}

There is currently great interest in the development of special-purpose quantum computing devices and methodologies that do not require full error correction and which are practical now. For example, D-Wave Systems produces commercial quantum annealers based on superconducting circuits that solve an important class of binary optimization problems \cite{JohnsonNat11}. However it is not known whether the D-Wave annealers can outperform conventional classical supercomputers \cite{BoixoNatPhys14,RonnowSci14}. An optical approach \cite{AaronsonTheorComp13} that solves an arguably less important problem---sampling from the distribution of bosons scattered by a unitary network---but which is likely capable of quantum speedup has also been investigated \cite{BroomeSci13,SpringSci13,TillmannNatPhotonics13}.  An approach called the single-excitation-subspace (SES) method, also based on supercondonducting circuits, has been proposed  
[\onlinecite{GellerMartinisEtalPRA15}].
Here computations are performed in the $n$-dimensional SES of a complete graph of $n$ qubits.
We call these examples {\it prethreshold}, referring to the threshold theorem of fault-tolerant quantum computation, because they do not require exceeding fidelity and qubit-number thresholds before being applicable.

\begin{figure}
\includegraphics[width=6.0cm]{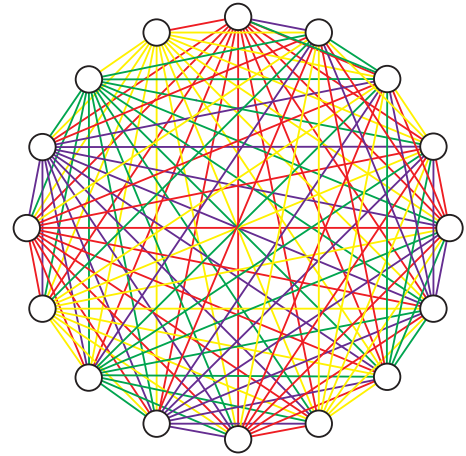} 
\caption{(Color online) Complete graph with $n \! = \! 16$. The vertices (open circles) are qubits and the edges (colored lines) are tunable couplers.}
\label{CG16 figure}
\end{figure} 

\begin{figure}
{\vskip 0.2in}
\includegraphics[width=6cm]{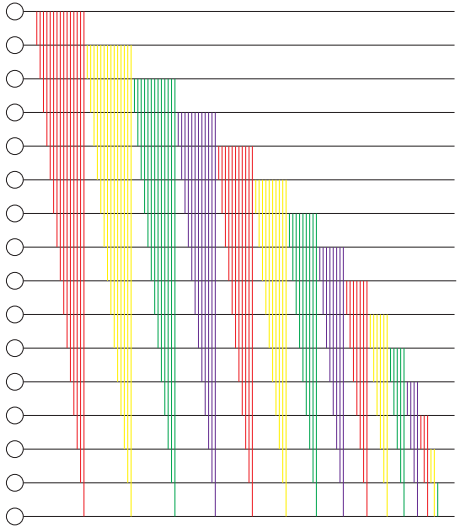} 
\caption{(Color online) Possible layout for the 16-qubit chip.}
\label{CG16 layout figure}
\end{figure}

A quantum computer chip implementing the SES method consists of a fully connected array of superconducting qubits with tunable frequencies and tunable pairwise $\sigma^x \! \otimes \sigma^x$ couplings; an abstract representation is given in Fig.~\ref{CG16 figure}. It works by operating in a subspace of the full $2^n$-dimensional Hilbert space where the Hamiltonian can be directly programmed. This programmability eliminates the need to decompose operations into elementary one- and two-qubit gates, enabling larger computations to be performed within the available coherence time. The price for this high degree of controllability is that the approach is not scalable. However, a technically unscalable quantum computer is still useful for prethreshold quantum computation and might even be able to achieve speedup relative to a classical supercomputer for certain tasks. The SES approach trades physical qubits and high connectivity for, in effect, longer coherence. This is a sensible trade for quantum computing architectures such as superconducting circuits, whose largest prethreshold problem sizes are limited by coherence time, not by the difficulty of introducing additional qubits. A realistic chip layout that provides space for the coupler circuits and avoids the crossovers of Fig.~\ref{CG16 figure} is shown in Fig.~\ref{CG16 layout figure}. 

\section{CONTENT OF THIS PAPER}

A significant restriction of the SES method presented in Ref.~\cite{GellerMartinisEtalPRA15} is that the Hamiltonian programmed into the hardware is real and symmetric, whereas the most general Hamiltonian is complex Hermitian. If a target operation has the form $e^{-i A},$ where $A$ is a known real symmetric generator matrix, then the unitary can be implemented in one step. This is the case when the unitary is {\it symmetric} ($U = U^\top$) and is reviewed in Sec.~\ref{single-step implementation of symmetric unitaries subsection}. In that section we also provide an improved procedure for constructing the time-optimal SES Hamiltonian ${\cal H}$ corresponding to a given generator $A$.

However, a general element of the unitary group ${\rm U}(n)$ has the form $e^{-i M}$ with $M$ complex Hermitian. This is the {\it nonsymmetric} unitary case ($U \neq U^\top$) discussed in Sec.~\ref{three-step implementation of nonsymmetric unitaries subsection}. We show there that any nonsymmetric element $U \in {\rm U}(n)$ can be implemented in three steps, for any $n$. 

Applications of these techniques are given in Sec.~\ref{applications section}. In Sec.~\ref{hamiltonian simulation subsection} we show how to simulate time-independent but otherwise arbitrary $n\!\times\!n$ complex Hamiltonians with an SES chip in three steps. In Sec.~\ref{state preparation subsection} we show how to prepare pure but otherwise arbitrary SES states in three steps. And in Sec.~\ref{expectation values subsection} we explain how to compute expectation values of arbitrary $n\!\times\!n$ Hermitian observables.

\section{SES IMPLEMENTATION OF UNITARY OPERATORS}

\subsection{Single-step implementation of symmetric unitaries}
\label{single-step implementation of symmetric unitaries subsection}

The basic single-step operation in SES quantum computing is the implementation of symmetric unitaries of the form $U=e^{-iA}$, with $A$ real and symmetric \cite{GellerMartinisEtalPRA15}. Therefore, a standard task in SES algorithm design and implementation is the construction of an optimal protocol---an SES Hamiltonian ${\cal H}$ and evolution time $t_{\rm qc}$---to implement that unitary. We assume here that the generator matrix $A$ is {\it known}; if it is not then the classical overhead for obtaining $A$ from $U$ must be included in the quantum runtime. (We also note that the generator $A=i \log U$ is not unique.) The optimal protocol for implementing a symmetric unitary depends on the functionality assumed of the chip, especially of the tunable coupler circuits. Here we assume that the experimentally controlled SES Hamiltonian can be written, apart from an additive constant, as
\begin{equation}
{\cal H} = g_{\rm max} K 
\ \ \ {\rm with} \ \ \ -1 \le K_{i i'} \le 1,
\label{standard form}
\end{equation}
which we call the {\it standard form}. In this case we are assuming that the couplings can be tuned continuously between $-g_{\rm max}$ and $g_{\rm max}$, and that the qubit frequencies can be varied within a window of width $2 g_{\rm max}$ about some parking frequency. Because we are free to change the overall phase of an SES state, we write the symmetric unitary as
\begin{equation}
U = e^{-i(A-cI)} e^{-ic}, 
\label{shited symmetric unitary}
\end{equation}
where $I$ is the $n\!\times\!n$ identity matrix, 
and then ignore the global phase $e^{-ic}$. The value of $c$ is chosen to minimize the evolution time $t_{\rm qc}$, which is proportional to the angle
\begin{equation}
\theta_{\!A} \equiv \max_{i i'} |A_{ii'} - c \delta_{ii'}|.
\label{angle}
\end{equation}
The $K$ matrix in (\ref{standard form}) is then given by
\begin{equation}
K = \frac{A-cI}{\theta_{\!A}},
\label{standard form K}
\end{equation}
and the evolution time is
\begin{equation}
t_{\rm qc}= \frac{\hbar \theta_{\!A}}{g_{\rm max}}. 
\label{standard form evolution time}
\end{equation}
Note that $\theta_{\!A}$ is not bounded by $2\pi$ and can become arbitrarily large. The global phase angle that minimizes
$\theta_{\!A}$ is
\begin{equation}
c = \frac{\min_{i} A_{ii}+\max_{i} A_{ii}}{2},
\end{equation}
which is proved below. Although we have assumed that the SES Hamiltonian ${\cal H} = g_{\rm max} K$ is abruptly switched on for a time $t_{\rm qc}$ before being abruptly switched off---which is the fastest protocol---any SES Hamiltonian of the form ${\cal H} = g(t) K$ such that $\int (g/\hbar) \, dt = \theta_{\!A}$ may be used instead.

To minimize (\ref{angle}) over $c$ we consider two cases: In the first case $\max_{ii'} |A_{ii'}|$ occurs for an {\it off-diagonal} element of $A$, in which case the minimum value of $\theta_{\!A}$ is independent of $c$ (because $c$ only affects the diagonal elements of the shifted matrix $A-c I$). Therefore we only need to consider the second case where $\max_{ii'} |A_{ii'}|$ occurs for a {\it diagonal} element. The diagonal elements consist of points 
\begin{equation}
x \in \big\lbrace A_{11}, A_{22}, \cdots , A_{nn} \big\rbrace
\label{diagonal element points}
\end{equation}
on the real number line, bounded between $\min_{i} A_{ii}$ and $\max_{i} A_{ii}$. Placing $c$ at the midpoint of the smallest region containing all the points in (\ref{diagonal element points}) minimizes the largest distance $|A_{ii} - c|.$ 

\subsection{Three-step implementation of nonsymmetric unitaries: ABA decomposition}
\label{three-step implementation of nonsymmetric unitaries subsection}

Our protocol relies on the matrix decomposition 
\begin{equation}
U = O_1 e^{-iD} O_2^\top,  
\label{KAK decomposition}
\end{equation}
where $D$ is a real diagonal matrix and the $O_i \in O(n)$ are real orthogonal matrices. This identity follows from the $KAK$ decomposition of the Lie group ${\rm U}(n) \!$ \cite{KnappLieGroups}. To obtain the $O_i$ and $D$ from $U$, we first compute
\begin{equation}
\chi \equiv U U^\top = O_1 e^{-2iD} O_1^\top,
\end{equation}
which is both symmetric and unitary. The real and imaginary parts of $\chi$ are also separately symmetric. Then the unitarity condition
\begin{equation}
\big({\rm Re}\, \chi - i \, {\rm Im} \, \chi\big)
\big({\rm Re}\, \chi + i \, {\rm Im} \, \chi\big) = I\end{equation}
shows that ${\rm Re}\, \chi$ and ${\rm Im}\, \chi$ commute and can be simultaneously diagonalized. $O_1$ is determined by a Schur decomposition of  ${\rm Re}\, \chi$, which always produces a real $O_1$ (unlike the decomposition of $\chi$ itself). Then $e^{-2iD}$ and $O_2$ are obtained from $O_1^\top \chi O_1$ and $U^\top O_1 e^{iD} \! ,$ respectively. 

The three-step implementation for a nonsymmetric $U \in {\rm U}(n)$ follows from the identity 
\begin{equation}
U = e^{-iA} e^{-iB} e^{iA},
\label{ABA decomposition}
\end{equation}
which we call the {\it ABA decomposition}. Here $A$ and $B$ are real symmetric $n \! \times \! n$ matrices. To derive (\ref{ABA decomposition}) we express the target unitary in the spectral form $U = V e^{-i \Lambda} V^\dagger,$ where $V$ is complex unitary and $\Lambda$ is real and diagonal. Decomposing $V$ using (\ref{KAK decomposition}) we have
\begin{eqnarray}
U &=& O_1 e^{-iD} O_2^\top e^{-i \Lambda} O_2 \, e^{iD} O_1^\top, 
\nonumber \\
&=& e^{-i O_1 D O_1^\top} (O_1 O_2^\top) \, e^{-i \Lambda} \, (O_1 O_2^\top)^{\! \top} e^{i O_1 D O_1^\top},
\end{eqnarray}
which leads to (\ref{ABA decomposition}) with generators
\begin{eqnarray}
A &=& O_1 D O_1^\top , \label{A formula} \\
B &=& O_1 O_2^\top \Lambda O_2  O_1^\top,
\label{B formula}
\end{eqnarray}
which are both real and symmetric. The classical runtime to obtain $A$ and $B$ is about 
\begin{equation}
1.4 \! \times \! n^{2.3} \, {\rm \mu s}
\label{ABA decomposition classical preprocessing time}
\end{equation}
on a laptop computer \cite{computationNote}.  The quantum runtime to implement a nonsymmetric unitary is
\begin{equation}
t_{\rm qc}= \frac{\hbar (2\theta_{\!A}+\theta_{\!B})   }{g_{\rm max}},
\label{nonsymmetric unitary runtime}
\end{equation}
with $\theta$ defined in (\ref{angle}). The generator matrices $A$ and $B$ in (\ref{ABA decomposition}) are not unique.

The ABA decomposition allows for the possibility of implementing highly complex operations in three steps. But this does not imply that an entire algorithm, compiled into a single unitary, can be implemented in constant time, because the compiled unitary might not be known a priori, and there is classical overhead (\ref{ABA decomposition classical preprocessing time}) for computing $A$ and $B$. More importantly, evaluating $A$ and $B$ for an entire algorithm would presumably be prohibitive when one is attempting to outperform classical computers. Furthermore, algorithms might include measurement steps that cannot be postponed to the end.

\section{APPLICATIONS}
\label{applications section}

\subsection{Hamiltonian simulation}
\label{hamiltonian simulation subsection}

A useful application of (\ref{ABA decomposition}) is to $U = e^{-iHt/\hbar},$ where $H$ is a given complex Hamiltonian. In this case we have 
\begin{equation}
e^{-iHt/\hbar} = e^{-iA} e^{-iB} e^{iA},
\label{complex Hamiltonian simulation}
\end{equation}
with $A$ and $B$ given by (\ref{A formula}) and
(\ref{B formula}), where $\Lambda$ is a diagonal matrix containing $t/\hbar$ times the spectrum of $H$. This enables the fast simulation of any time-independent Hamiltonian with an SES chip \cite{timeIndependentNote}.

\subsection{SES pure state preparation in 3 steps}
\label{state preparation subsection}

In some cases it is possible to compile an entire algorithm down to only three steps. As an example we give an algorithm for preparing any (normalized) pure SES state of the form
\begin{equation}
| \psi\rangle = \sum_{i=1}^n a_i \, | i ) , \ \  \  \  a_i = |a_i|e^{i \theta_i}, \ \ \ \ 0 \le \theta_i < 2\pi.
\label{target SES state}
\end{equation}
Here $| i ) \equiv |0 \cdots 1_i  \cdots 0\rangle$ is the $i$th SES {\it basis state} of the $n$-qubit processor. We proceed by giving a protocol with linear depth that is practical for small $n$, which is then subsequently compiled down to three steps. 

We start with the basis state $|1)$, which is prepared from the system ground state $|00\cdots0\rangle$ by a microwave pulse, and then apply the standard-form SES Hamiltonian ${\cal H} = g_{\rm max} K_{\rm star}$ for a time $t_{\rm qc} = \pi \hbar /\sqrt{n} g_{\rm max},$ with
\begin{equation}
K_{\rm star} \equiv
\begin{pmatrix}
1 & \frac{1}{2} & \frac{1}{2} & \cdots & \frac{1}{2} \\
\frac{1}{2} & 0 & 0 & \cdots & 0 \\
\frac{1}{2} & 0 & 0 & \cdots & 0 \\
\vdots & \vdots & \vdots  & \ddots & \vdots \\
 \frac{1}{2} & 0 & 0 & \cdots & 0 \\
\end{pmatrix}
\label{K star definition}
\end{equation}
the adjacency matrix for a star graph with qubit 1 at the center (see Sec.~IIIA of Ref.~[\onlinecite{GellerMartinisEtalPRA15}]). 
This produces the {\it uniform} state
\begin{equation}
\big|{\rm unif} \big\rangle \equiv \frac{| 1 ) +  | 2) + \cdots + | n)}{\sqrt{n}},
\label{uniform state}
\end{equation}
apart from a phase.

\begin{figure}
\includegraphics[width=8.0cm]{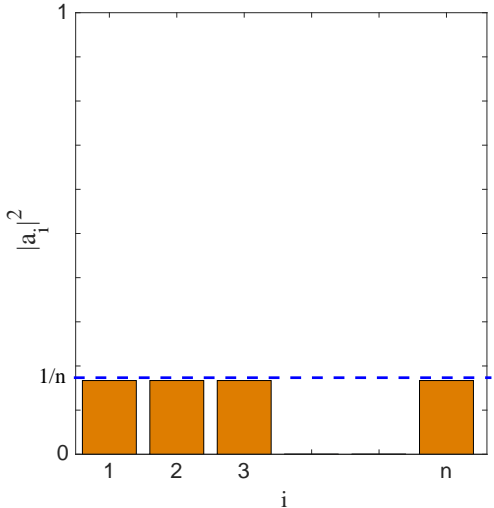} 
\caption{(color online) Occupation probabilities for a uniform weight state. Phases of the probability amplitudes $a_i$ are not represented in this figure.}
\label{uniform weight figure}
\end{figure}

If the occupation probabilities in the target state (\ref{target SES state}) are uniform,
\begin{equation}
|a_i|^2 = \frac{1}{n},
\label{uniform weight state}
\end{equation}
we call it a {\it uniform weight} state and represent it by the bar graph in Fig.~\ref{uniform weight figure}. In this case we would apply the diagonal Hamiltonian ${\cal H} = g_{\rm max} K$, where
\begin{equation}
K=-
\begin{pmatrix}
\frac{\theta_1}{2\pi} & 0 & \cdots & 0 \\
0 &\frac{\theta_2}{2\pi}  & \cdots & 0 \\
\vdots & \vdots  & \ddots & \vdots \\
0 &  0 & \cdots & \frac{\theta_n}{2\pi} \\
\end{pmatrix}
\label{diagonal phase shifting Hamiltonian}
\end{equation}
to the uniform state $|{\rm unif}\rangle$ for a time $t_{\rm qc} = 2\pi \hbar /g_{\rm max},$ which gives the desired target.

Typically the target is not a uniform weight state, as  represented in Fig.~\ref{initial weight figure}. In this case we use the solution 
\begin{equation}
|{\rm unif}\rangle = W_{\rm diag} \, (U_{\rm swap} U_{\rm diag})_M \! \cdots  (U_{\rm swap} U_{\rm diag})_1 |\psi\rangle
\label{inverse problem}
\end{equation}
to the inverse problem of constructing the uniform state $|{\rm unif}\rangle$ from the target \cite{LawPRL96,HofheinzNat09}. Each of the $M$ steps in (\ref{inverse problem}) consists of a pair of operations $U_{\rm diag}$ and $U_{\rm swap}$ that move weight between a pair of components. After $M = O(n)$ steps a uniform weight state is created. The final operation $W_{\rm diag}$ shifts the phases of the uniform weight state to that of (\ref{uniform state}). The first step is:
\begin{figure}
\includegraphics[width=8.0cm]{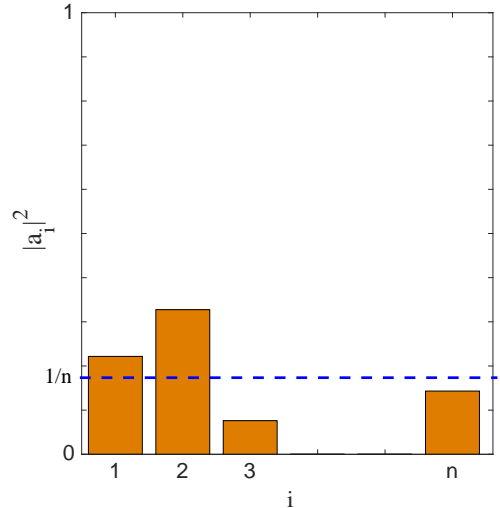} 
\caption{(color online) Occupation probabilities for a typical target state. In this example the components of maximum and minimum weight have indices $i_{\rm max} \! = \! 2$ and $i_{\rm min} \! = \! 3$.}
\label{initial weight figure}
\end{figure} 
\begin{enumerate}

\item

Find the components $i_{\rm min}$ and $i_{\rm max}$ with the smallest and largest weights, respectively (if not unique, any solution is sufficient). These satisfy\begin{equation}
|a_{i_{\rm min}}|^2 \le  \frac{1}{n} \le |a_{i_{\rm max}}|^2 ,
\label{weight inequalities}
\end{equation}
excluding the case where both $\le$ signs are identities (which would violate the assumption that the target is nonuniform). Therefore $|a_{i_{\rm min}}|^2 <  |a_{i_{\rm max}}|^2 \! .$ 

\item

Perform a phase shift $U_{\rm diag}= e^{-i {\cal H} t_{\rm qc}/\hbar}$ that brings the probability amplitudes $a_{i_{\rm min}}$ and $a_{i_{\rm max}}$ to the form $a_{i_{\rm min}} = |a_{i_{\rm min}}|$ and $a_{i_{\rm max}} =i |a_{i_{\rm max}}|,$ with $|a_{i_{\rm min}}| <  |a_{i_{\rm max}}|$. Apply SES Hamiltonian (\ref{standard form}), where $K$ is a diagonal matrix with $K_{i_{\rm min},i_{\rm min}}= \theta_{i_{\rm min}}/3\pi$ and  $K_{i_{\rm max},i_{\rm max}}=(\theta_{i_{\rm max}}/3\pi)- \frac{1}{6}$, the other elements zero, and $t_{\rm qc} = 3 \pi \hbar /g_{\rm max}.$ This phase shift  is necessary to prepare the state for the next operation.

\item 

Implement a partial {\sf iSWAP} $U_{\rm swap}= e^{-i {\cal H} t_{\rm qc} /\hbar }$ from component $i_{\rm max}$ to $i_{\rm min}$ to bring the weight of $i_{\rm min}$ to the uniform value, 
\begin{equation}
|a_{i_{\rm min}}|^2 \rightarrow \frac{1}{n},
\end{equation}
and leaving component $i_{\rm max}$ with weight
\begin{equation}
|a_{i_{\rm max}}|^2 \rightarrow |a_{i_{\rm min}}|^2 + |a_{i_{\rm max}}|^2 - \frac{1}{n}.
\end{equation}
Apply SES Hamiltonian (\ref{standard form}) with $K_{i_{\rm min},i_{\rm max}} \! = \! K_{i_{\rm max},i_{\rm min}} \! = \! 1$ and all other elements zero, and $t_{\rm qc} = \varphi \hbar /g_{\rm max}$ with $\varphi$ given by
\begin{equation}
|a_{i_{\rm min}}| \cos\varphi + |a_{i_{\rm max}}| \sin \varphi = \sqrt{1/n}.
\end{equation}
There is always a solution with $0 \! < \! \varphi \! <  \! \pi/2$. 
\end{enumerate}
This completes the first step.

If after the first step $(U_{\rm swap} U_{\rm diag})_1 |\psi\rangle$ is a uniform weight state, it can be written in the form
\begin{equation}
\frac{e^{i \alpha_1} | 1 ) +  e^{i \alpha_2} | 2) + \cdots + e^{i \alpha_n}| n)}{\sqrt{n}},
\end{equation}
and we apply the final operation $W_{\rm diag} = e^{-i {\cal H} t_{\rm qc}/\hbar}$ to produce (\ref{uniform state}). Here we use SES Hamiltonian (\ref{standard form}) with
\begin{equation}
K=
\begin{pmatrix}
\frac{\alpha_1}{2\pi} & 0 & \cdots & 0 \\
0 &\frac{\alpha_2}{2\pi}  & \cdots & 0 \\
\vdots & \vdots  & \ddots & \vdots \\
0 &  0 & \cdots & \frac{\alpha_n}{2\pi} \\
\end{pmatrix}
\label{Hamiltonian for Vdiag}
\end{equation}
and $t_{\rm qc} = 2\pi \hbar /g_{\rm max}.$ If $(U_{\rm swap} U_{\rm diag})_1 |\psi\rangle$ is not a uniform weight state, we again find the minimum and maximum weight components $i_{\rm min}$ and $i_{\rm max}$, and follow the above protocol to generate $(U_{\rm swap} U_{\rm diag})_2 (U_{\rm swap} U_{\rm diag})_1  |\psi\rangle.$ The procedure is repeated until 
\begin{equation}
(U_{\rm swap} U_{\rm diag})_M \cdots (U_{\rm swap} U_{\rm diag})_2 (U_{\rm swap} U_{\rm diag})_1  |\psi\rangle
\end{equation}
is a uniform weight state, after which $W_{\rm diag}$ is applied. The number of iterations required satisfies
\begin{equation}
M \le n-1.
\label{M bound}
\end{equation}
This completes the solution to the inverse problem (\ref{inverse problem}). 

We now use (\ref{inverse problem}) to obtain
\begin{equation}
|\psi\rangle =  (U_{\rm diag}^\dagger U_{\rm swap}^\dagger )_1\cdots (U_{\rm diag}^\dagger U_{\rm swap}^\dagger )_M   \, W_{\rm diag}^\dagger |{\rm unif}\rangle,
\label{forward problem}
\end{equation}
which solves the general state-preparation problem in $O(n)$ steps. Hermitian conjugations are implemented by changing the signs of the $K$ matrices given above. The protocol given in (\ref{forward problem}) is, by itself, practical for small $n$.

The complete state preparation operation can be summarized as 
\begin{equation}
|\psi \rangle  = U \, |1),
\end{equation}
where
\begin{equation}
U \equiv  (U_{\rm diag}^\dagger U_{\rm swap}^\dagger )_1\cdots (U_{\rm diag}^\dagger U_{\rm swap}^\dagger )_M   \, W_{\rm diag}^\dagger
\, e^{-i \frac{\pi}{\sqrt{n}} K_{\rm star}} 
\label{compiled U definition}
\end{equation}
is the {\it compiled} unitary of the state-preparation algorithm.  The three-step state preparation protocol uses the ABA decomposition to implement (\ref{compiled U definition}). The total state preparation time, not including the $|1)$ state initialization time, is given in (\ref{nonsymmetric unitary runtime}).

For example, suppose we wish to prepare the randomly chosen target
\begin{eqnarray}
&|\psi\rangle \!=\! 0.4829 \, |1) + (-0.5478\!-\!0.0575i)\, |2) + (0.1142\!+\!0.2387i) \, |3) & \nonumber \\ 
&+ (0.4095\!+\!0.2400i) \, |4) + (-0.3215\!+\!0.2545i) \, |5),&
\label{target example}
\end{eqnarray}
in the $n\!=\!5$ graph, where for convenience the first component has been chosen to be real. Following the state-preparation protocol leads to the compiled unitary
\begin{widetext}
\begin{equation}
U=
\begin{pmatrix}
0.4829  & 0.4499 - 0.0158i &  0.4499 - 0.0158i  & 0.4478 - 0.0133i  & 0.3984 + 0.0450i \\
  -0.5478 - 0.0575i &  0.5855 - 0.4153i  & 0.1778 - 0.0249i & -0.1305 + 0.2703i & -0.0855 + 0.2273i \\
   0.1142 + 0.2387i  & 0.4664 + 0.0700i & -0.7862 - 0.2582i &  0.0910 - 0.0284i  & 0.1145 - 0.0222i \\
   0.4095 + 0.2400i  & 0.0841 - 0.1271i &  0.1471 - 0.1492i  &-0.7941 + 0.1818i  & 0.1471 - 0.1492i \\
  -0.3215 + 0.2545i &  0.1071 + 0.1577i  & 0.1071 + 0.1577i  & 0.1071 + 0.1580i  & 0.1399 - 0.8386i \\
\end{pmatrix},
\label{compiled U example}
\end{equation}
\end{widetext}
up to a phase factor. The first column of (\ref{compiled U example}) is the target state. The ABA decomposition (\ref{ABA decomposition}) then leads to
\begin{equation}
A\!=\!
\begin{pmatrix}
 -1.1145  &  0.1981 &   0.3247  & -0.0776  & -0.1888 \\
    0.1981 &  -2.6988  &  0.0219  & -0.2069  & -0.0249 \\
    0.3247  &  0.0219  & -1.9798  & -0.5623  &  0.1052 \\
   -0.0776  & -0.2069  & -0.5623   &-0.5291 &  -0.0747 \\
   -0.1888  & -0.0249  &  0.1052  & -0.0747  & -1.7104 \\
\end{pmatrix} 
\end{equation}
and
\begin{equation}
B\!=\!
\begin{pmatrix}
-3.0826  &  1.8972 &   0.3983  &  0.8753  &  0.5934 \\
    1.8972  & -3.7784  &  0.5761  &  0.3537  &  0.5581 \\
    0.3983  &  0.5761  & -3.2370  &  0.1664  &  0.2327 \\
    0.8753  &  0.3537  &  0.1664 &  -2.6191  &  0.1488 \\
    0.5934  &  0.5581  &  0.2327 &   0.1488  & -4.6171 \\
\end{pmatrix}.
\end{equation}
The associated $K$ matrices and evolution times are determined from the procedure given in Sec.~\ref{single-step implementation of symmetric unitaries subsection}:
\begin{eqnarray}
K_A&\!=\!&
\begin{pmatrix}
0.4604   & 0.1826  &  0.2993  & -0.0715 &  -0.1741 \\
    0.1826  & -1 &   0.0202  & -0.1907  & -0.0229 \\
    0.2993  &  0.0202  & -0.3373  & -0.5183  &  0.0970 \\
   -0.0715  & -0.1907  & -0.5183   & 1  &-0.0689 \\
   -0.1741  & -0.0229  &  0.0970  & -0.0689  & -0.0889 \\
   \end{pmatrix} , \nonumber \\  
\theta_A &=&  1.0848,
\end{eqnarray}
and
\begin{eqnarray}
K_B&\!=\!&
\begin{pmatrix}
 0.2822   & 1 &  0.2100  &  0.4614  &  0.3128 \\
    1 & -0.0845 &   0.3037  &  0.1864  &  0.2942 \\
    0.2100  &  0.3037  &  0.2009  &  0.0877  &  0.1226 \\
    0.4614  &  0.1864  &  0.0877  & 0.5266  &  0.0785 \\
    0.3128  &  0.2942   & 0.1226   & 0.0785  & -0.5266 \\
\end{pmatrix} , \nonumber \\  
\theta_B &=&  1.8972.
\end{eqnarray}
The total state preparation time, not counting the $|1)$ state initialization, is given by (\ref{nonsymmetric unitary runtime}).  This is about $13 \, {\rm ns}$ for the target state (\ref{target example}) in an SES chip with $g_{\rm max}/2\pi = 50 \, {\rm MHz}$. 

Although state preparation is implemented in three steps for any $n$, the runtime does have a weak $n$-dependence, because $\theta_{\!A}$ and $\theta_{\!B}$ do. Averaged over random targets we find that
\begin{equation} 
\overline{ 2 \theta_{\!A} + \theta_{\!B} } \approx 4.0 \! \times \! n^{0.06} \! .
\end{equation}

For small $n$, either the linear-depth protocol (\ref{forward problem}) or the three-step protocol based on (\ref{compiled U definition}) can be used. However for large $n$, only the three-step protocol is practical.

\subsection{Computation of expectation values}
\label{expectation values subsection}

Finally, we show how to compute the expectation value
\begin{equation} 
\langle O \rangle \equiv {\rm Tr} (\rho O)
\end{equation}
of any $n\!\times\!n$ Hermitian observable $O$, by implementing the protocol of Reck {\it et al.}~\cite{ReckPRL94}. Here $\rho$ is any pure or mixed SES state provided as an input to the procedure.

Standard readout of an SES processor consists of the simultaneous measurement of each qubit in the diagonal basis. The SES condition means that a single qubit will be found in the state $|1\rangle$, with the remaining $n-1$ qubits in $|0\rangle.$ Let $i$ be the qubit observed in it's excited state. The probability of observing the excitation in qubit $i$ is $p_i = (i|\rho|i)$. Therefore, if we have access to multiple copies of $\rho$ we can repeat the readout $N$ times to obtain estimates of the occupation probabilities $p_i$ with sampling errors no larger than $(2\sqrt{N})^{-1}$.
 
To compute $\langle O \rangle$, perform a (classical) spectral decomposition to a unitary $V$ containing the eigenvectors of $O$ as columns, and a real diagonal matrix $D$: $O=VDV^\dagger$. Then we have
\begin{equation}
\langle O \rangle = {\rm Tr} (\rho VDV^\dagger)={\rm Tr} (\rho^\prime D), 
\end{equation}
where
\begin{equation}
\rho^\prime\equiv V^\dagger \rho V.
\end{equation}
Therefore we can compute $\langle O \rangle$ by applying the unitary operator $V^\dagger$ using the ABA decomposition, measuring the resulting occupation probabilities, which we denote by $p_i^{\, \scriptscriptstyle[V^\dagger \!]}$ to indicate the application of $V^\dagger$, and then classically evaluating the quantity
\begin{equation}
\langle O \rangle = \sum_{i=1}^n D_{ii} \, p_i^{\, \scriptscriptstyle[V^\dagger \!]}.
\end{equation}

\section{CONCLUSIONS}

In this work we have extended the SES method of Ref.~\cite{GellerMartinisEtalPRA15} to include a three-step implementation of arbitrary $n \times n$ unitaries. The fast state preparation protocol of Sec.~\ref{state preparation subsection} should be especially useful for practical quantum computing applications.

\acknowledgements

This work was supported by the US National Science Foundation under CDI grant DMR-1029764. It is a pleasure to thank Emmanuel Donate and Timothy Steele for their contributions during the early stages of this work.

\bibliography{../../bibliographies/algorithms,../../bibliographies/dwave,../../bibliographies/control,../../bibliographies/error_correction,../../bibliographies/general,../../bibliographies/group,../../bibliographies/ions,../../bibliographies/math,../../bibliographies/nmr,../../bibliographies/optics,../../bibliographies/simulation,../../bibliographies/superconductors,../../bibliographies/surface_code,endnotes}

\begin{thebibliography}{14}
\bibitem [{tim()}]{timeIndependentNote}%
  \BibitemOpen
  \href@noop {} {}\bibinfo {note} {In principle it is possible to simulate
  time-dependent complex Hamiltonians as well. However this would require
  decomposing the evolution into a sequence of short time steps such that
  $H(t)$ is approximately constant within each time step, and applying the ABA
  decomposition at each time step. Given the classical overhead required at
  each step, this does not seem useful for outperforming classical
  computers.}\BibitemShut {Stop}%
\end{thebibliography}

\end{document}